\documentclass{article}
\usepackage{spconf,amsmath,graphicx}
\usepackage{enumitem}
\usepackage{listings}
\usepackage{xcolor}
\usepackage{textcomp}
\usepackage[T1]{fontenc}

\definecolor{codegreen}{rgb}{0,0.6,0}
\definecolor{codegray}{rgb}{0.5,0.5,0.5}
\definecolor{codepurple}{rgb}{0.58,0,0.82}
\definecolor{backcolour}{rgb}{0.95,0.95,0.92}

\lstdefinestyle{mystyle}{
    backgroundcolor=\color{backcolour},   
    commentstyle=\color{codegreen},
    keywordstyle=\color{magenta},
    numberstyle=\tiny\color{codegray},
    stringstyle=\color{codepurple},
    basicstyle=\ttfamily\footnotesize,
    breakatwhitespace=false,         
    breaklines=true,                 
    captionpos=b,                    
    keepspaces=true,                 
    numbers=left,                    
    numbersep=5pt,                  
    showspaces=false,                
    showstringspaces=false,
    showtabs=false,                  
    tabsize=2
}

\title{Spatial Scaper: A library to simulate and augment soundscapes for sound event localization and detection in realistic rooms}
\name{Iran R. Roman$^{1*}$ \quad Christopher Ick$^{1*}$ \quad Sivan Ding$^{1}$ \quad Adrian S. Roman$^{2}$ \quad Brian McFee$^{1}$ \quad Juan P. Bello$^{1}$}

\address{$^{1}$ Music and Audio Research Laboratory, New York University, New York, USA \\ $^{2}$ Viterbi School of Engineering, University of Southern California, California, USA\\
{\normalsize $^*$Equal contribution}
}

\begin{document}
\maketitle
\begin{abstract}
Sound event localization and detection (SELD) is an important task in machine listening.
Major advancements rely on simulated data with sound events in specific rooms and strong spatio-temporal labels.
% This data comes with a high human-labor cost, simulated SELD data has also been used.
SELD data is simulated by convolving spatialy-localized room impulse responses (RIRs) with sound waveforms to place sound events in a soundscape.
However, RIRs require manual collection in specific rooms.
We present \texttt{SpatialScaper}, a library for SELD data simulation and augmentation.
Compared to existing tools, \texttt{SpatialScaper} emulates virtual rooms via parameters such as size and wall absorption.
This allows for parameterized placement (including movement) of foreground and background sound sources.
\texttt{SpatialScaper} also includes data augmentation pipelines that can be applied to existing SELD data. 
As a case study, we use \texttt{SpatialScaper} to add rooms to the DCASE SELD data.
Training a model with our data led to progressive performance improves as a direct function of acoustic diversity.
These results show that \texttt{SpatialScaper} is valuable to train robust SELD models.
\end{abstract}%
\section{Introduction}
\label{sec:intro}

Sound event localization and detection (SELD) consists of two subtasks: localizing sound sources and determining their category (i.e. {\it music} vs {\it dog bark}) \cite{grumiaux2022survey}.
SELD is relevant for assistive technologies that improve the lifestyle and safety of low vision and audition individuals \cite{pandya2021ambient}.

Training models requires strongly-labeled data collected in rooms using microphone arrays \cite{politis2020overview}.
Curating data is labor-intensive and thus only a handful datasets exist \cite{shimada2023starss23,lollmann2018locata}.
An alternative is to simulate data using room impulse responses (RIRs) \cite{adavanne2019multi}. 
Since an RIR's location is known, convolving it with a sound waveform results in a soundscape with an event whose location, class, and start/end time are perfectly known. 
This method allows for data simulation at scale. 
However, this method still assumes RIRs collected in real rooms.

We present \texttt{SpatialScaper}, an open-source library for SELD data simulation and augmentation. 
Existing methods assume a RIR database collected in real rooms and using specific microphone hardware. 
\texttt{SpatialScaper}, in constrast, emulates virtual rooms of any size to synthesize RIRs using a microphone array of any shape.
This dramatically increases the range of acoustic diversity that \texttt{SpatialScaper} can simulate.
Moreover, \texttt{SpatialScaper} utilizes many databases of real RIRs, allowing for simulations in real and synthetic rooms. 
\texttt{SpatialScaper}'s API parametrizes a soundscape's most important variables, such as the room size, wall absorption, the audio files that function as background and foreground events, their location in the room, etc.
These can be user-defined or drawn from a distribution. 
Furthermore, \texttt{SpatialScaper} can also apply effects to individual sound events for data augmentation.
Moreover, it can augment existing SELD datasets using techniques known to improve SELD metrics \cite{wang2023four}.
We include a case study to showcase how using \texttt{SpatialScaper} leads to improved SELD model performance.
Our contributions:
\begin{enumerate}[nolistsep]
\item A library to simulate SELD data with strong labels by using both real and synthetic RIRs.\footnote{https://github.com/iranroman/SpatialScaper} 
%, with an API that for spatio-temporal control of sound events.
\item A study that highlights how increased acoustic diversity in SELD training data improves model performance.
\end{enumerate}

\section{Related work}
\label{sec:prior}
\begin{figure}
\includegraphics[width=8.5cm]{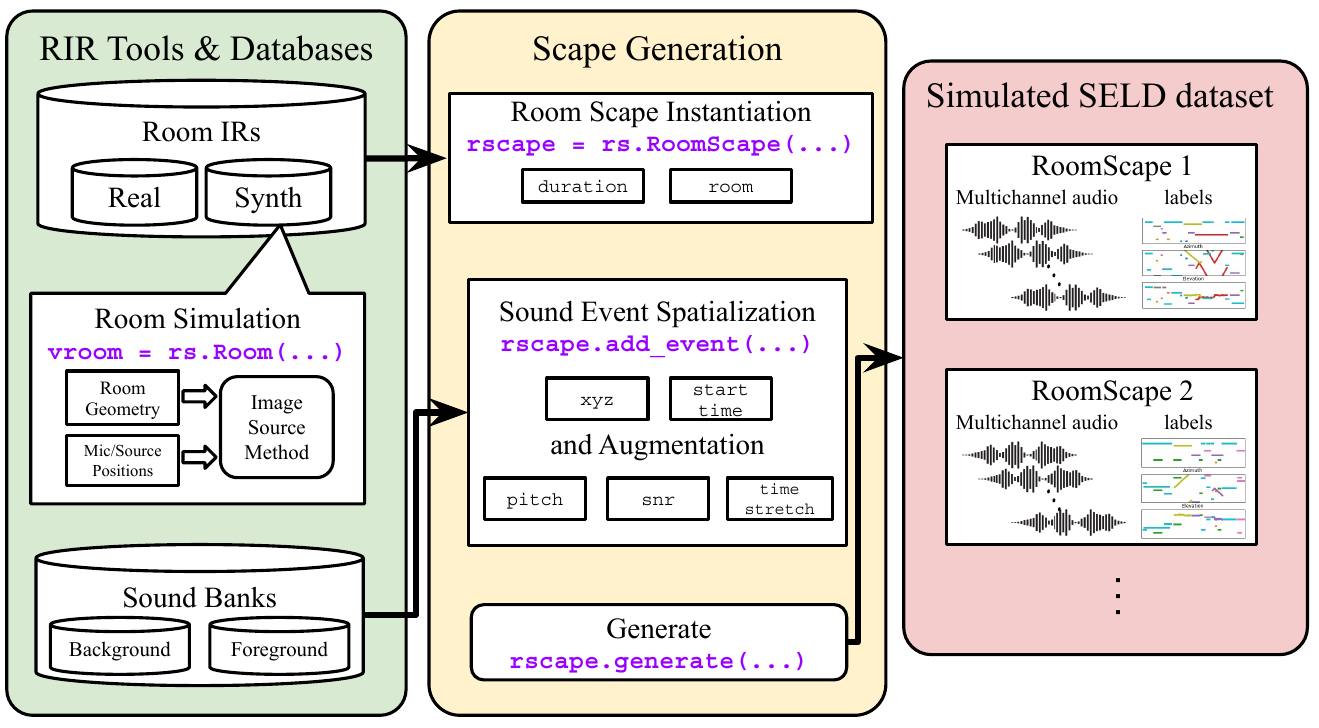}
\vspace{-10pt}
\caption{\texttt{SpatialScaper} data generation pipeline.}
\vspace{-15pt}
\label{fig:scaper}
\end{figure}

% While %joint task of 
While SELD is traditionally done via  traditional signal-processing \cite{param1}, modern approaches use deep neural networks \cite{grumiaux2022survey, politis2020overview}.
Major highlights over the past five years include: 1) the introduction of SELDnet \cite{politis2020overview}, a model that is updated every year to serve as a SELD baseline \cite{politis2020overview}, 2) the development of the muli-ACCDOA representation to detect multiple and overlapping sound sources even of the same class \cite{shimada2022multi}, 3) data augmentation techniques that can be applied to existing SELD datasets \cite{wang2023four}, and 4) robust SELD systems \cite{wang2023four, hu2022}.
These have been made possible in part due to SELD challenges, such as Detection and Classification of Acoustic Sound Events (DCASE) \cite{politis2020overview} or LOCalization And TrAcking (LOCATA) \cite{lollmann2018locata}, as well as the SELD datasets released for the 2019, 2020, and 2021 editions of DCASE consist of simulated data \cite{adavanne2019multi,politis2020dataset,politis2021dataset}.
In 2022 and 2023, the DCASE challenge added a relatively small dataset of strongly-annotated recordings of sound events produced by human actors in the real world known as STARSS \cite{shimada2023starss23}.

Several tools exist to simulate strongly-annotated soundscapes. 
One is \texttt{ambiscaper}, a library to simulate SELD data in the ambisonics spatial audio format \cite{ambiscaper}. 
Its limitations include the lack of dynamic sound sources (i.e. events that move), and no control of the distance between the microphone and the sound source, both of which are critical to develop ``complete'' SELD models \cite{grumiaux2022survey,kushwaha2023sound,liang2023reconstructing}. 
It is also limited by the fact that it only simulates the first order ambisonics (FOA) format.
Another code-base to simulate SELD data are scripts provided by the DCASE challenge organizers \cite{dcasegenerator}, which lack an API and thus offer limited possibilities to control simulation parameters.
Furthermore, this codebase uses a fixed number of rooms in its RIR database with FOA and 4ch tetrahedral microphone (which we refer to as ``MIC'') formats. 
Other RIR databases that utililze other microphone formats are available \cite{arni, metu}, but have not yet been used to generate SELD data.
Furthermore, recent SELD literature shows promising results by utilizing geometric room simulations to create synthetic RIRs for generating SELD data \cite{srivastava2023virtually, ickroomsim}.

One of the most cited libraries for single-channel soundscape generation is \texttt{scaper} \cite{scaper}.
However it only generates sound event detection (SED) data, lacking localization cues.
\cite{scaper}'s importance is rooted in its full parameterization of effects (like pitch shifting or time stretching) and temporal placement of sound events.
Parameters can be drawn from distributions to mitigate human bias data collection or simulation.
We build upon \texttt{scaper} to add spatial control of rooms, delivering the first library for parameterized SELD data generation 

\begin{figure}[t]
\centering
\begin{lstlisting}[language=Python, upquote=True]
import SpatialScaper as ss

# define a virtual room
vroom = ss.Room(
   dims = [5,3,2], decay = 0.8, mic_type = 'em32', 
   mic_loc = [2.5,2.5,0.5]) # mic_type could be 
   # MIC, a list of capsule coordinates, etc.

# create a room scaper instance
ssc = ss.SScaper(duration = 60, room = vroom, 
   fg_path = '/path/to/fg_events',
   bg_path = '/path/to/bg_events',
   ref_db = -50, # in dB
   )

# Add background noise
ssc.add_background(label = ('const', 'back'),
   source_file = ('choose', []),
   source_time = ('const', 0))
   )
                                   
# Add a moving sound event
ssc.add_event(label = ('choose', []),
   event_xyz = ('const', 
         [
            [4.0,0.1,0.2], [4.5,0.1,1.9]
         ] # inital and final position 
      )
   )
                     
# Genereate the audio and the annotation
ssc.generate(dest_path = '/path/out/rs1.wav')
\end{lstlisting}
  \caption{Instantiating a soundscape using a virtual room,  microphone, background noise, and a moving foreground event.}
  \label{gen_api}
\end{figure}

\vspace{-0.6cm}
\section{Spatial Scaper}

\texttt{SpatialScaper}'s primary use-case is creating soundscapes using virtual or real RIRs. 
In the case of a virtual room, the room size, microphone array, and sound decay are controllable.
Figure \ref{fig:scaper} gives an overview of the generation pipeline and \ref{gen_api} shows example code to simulate a soundscape.
The next subsections explain the API functionalities.

\subsection{Instantiating a room scape}
\texttt{SpatialScaper} integrates databases of RIRs and/or a virtual RIR simulation engine that uses the image source method implementation of \texttt{pyroomacoustics} \cite{pyroomacoustics}.
Line 4 in Figure \ref{gen_api} shows how a virtual room is defined.
Required parameters include the room dimensions, decay factor for sounds reflecting off of walls, and a microphone in a location.
Next, line 10 shows the instantiation of the soundscape.
Parameters include the duration of the output file, the room definition, and the paths to possible foreground and background sound files. 
While the example in Figure \ref{gen_api} uses a virtual room, the user could also specify one of the rooms in our curated database of RIRs (\texttt{METU} \cite{metu}, for example).

\subsection{Adding background noise}

Sustained background noise (i.e. AC or traffic) can be added to the room soundscape at a specific location with a constant SNR (usually a low value). 
Assuming that the \texttt{/path/to/bg\_events} contains files with sustained long noises, one or more can be selected and placed in the room soundscape (files will be looped throughout the track), using a \texttt{ref\_db} level.
Line 17 randomly selects a file from the path with background events and places it in a specific location. 
Alternatively, the background noise can be linearly added to all RIR channels to remove the localization effect. 

\subsection{Spatializing target events}

Placing a target event involves selecting a file, determining the time-point where it will start playing (both within itself and in the room soundscape), assigning it an SNR level, and determining its initial and final location in the room.
Effects such as random pitch-shift and time stretching may be applied. 
Line 23 in Figure \ref{gen_api} shows how a sound event is randomly selected to transverse the room between two xyz coordinates (specified by a list of two xyz coordinates).
By default, the sound event will move linearly (i.e. along the shortest path) throughout its duration. 
Users may also specify other possible trajectories (spline,  random walks, etc.).\footnote{Presently, free spatialization is only available in the virtual rooms.}
When spatializing events in a real room, the trajectory is adjusted to the nearest one possible given the RIR layout recorded in the room. 

\subsection{Triggering the room scape generation}

Finally, the soundscape generation. See line 32 in Figure \ref{gen_api} specifying the destination path.
The conversion between microphone and ambisonics formats is facilitated by \texttt{SpatialScaper}'s built-in ambisonics encoder, adapted and improved from the related \texttt{micarraylib} library \cite{roman2021micarraylib}. 

\subsection{Augmenting existing SELD recordings}

Given an existing SELD dataset, \texttt{SpatialScaper} can be used to augment using techniques like channel swapping, soundscape rotation, time-domain remixing, and random time-frequency masking \cite{wang2023four}. 
\texttt{SpatialScaper} generalizes these augmentations for any spherical array. 
Figure \ref{aug_api} shows an example augmentation pipeline. 
It assumes that the directory with the SELD dataset has a metadata subdirectory with csv files consistent with the DCASE SELD challenge format.
The API searches for all wav files with names that match the csv filenames to apply the augmentations. 

\begin{figure}
\centering
\begin{lstlisting}[language=Python, upquote=True]
ss.apply_augmentation(data_path='path/to/dataset',
                    aug_type = 'channel swapping')
# outputs path is 'path/to/dataset_swapped'

ss.apply_augmentation(data_path='/path/to/data',
                    aug_type = 'time freq mask')
# output path is 'path/to/dataset_tfmasked'
\end{lstlisting}
  \caption{Using \texttt{SpatialScaper} to augment a SELD dataset via the augmentations recently proposed by Wang et al. \cite{wang2023four}}
  \label{aug_api}
\end{figure}

\section{Case study: improving SELDnet}

\subsection{Model, training procedure, and metrics}

SELDnet is the baseline model that the organizers of the DCASE SELD challenge update every year \cite{politis2020overview}.
It is a convolutional recurrent neural network that can take either MIC or FOA  inputs. 
Its output is the multi-ACCDOA representation that can detect and classify multiple and overlapping sound sources, even of the same class \cite{shimada2022multi}.
We use the SELDnet that was released with the 2023 version of STARSS \cite{shimada2023starss23}. 

To replicate SELDnet's training procedure, one must use data that is both simulated and recorded in the real world. 
The simulated data comes from the ``DCASE 2022 SELD mixtures for baseline training'' \cite{politis_archontis_2022_6406873}, which we refer to as ``DCASE'' dataset.
The real data comes from the STARSS dataset \cite{shimada2023starss23}. 
SELDnet is trained with ``DCASE'' and the ``dev-train'' files in STARSS.
In its official implementation the model is cross-validated on ``dev-test'' files in STARSS, and is ``tested'' on the same ``dev-test'' files. 
This is done because the annotations for the STARSS evaluation files are not publicly available. To ensure that we cross-validate and test on annotated recordings carried out in separate rooms, we divide the STARSS ``dev-test'' by recording location. 
We use the rooms that were recorded in Finland (``dev-test-tau'') for cross-validation and the rooms that were recorded in Japan (``dev-test-sony'') for the final test. 
This ensures that the sounds and rooms used for the final evaluation are not seen during model development. 

We use the DCASE SELD metrics of location-dependent F-score (F) and error rate (ER) for classification, and localization error (LE) and recall (LR) \cite{politis2020overview}.
SELDnet shows optimal performance on FOA, so our experiments use that format. 

% \vspace{-0.3cm}
\subsection{Exp 1: Adding acoustic diversity to the training data}

The ``DCASE'' dataset consists of simulated SELD data using RIRs collected in nine different rooms (150 1min soundscapes per room). 
We use \texttt{SpatialScaper} to simulate more room soundscapes using real and synthetic RIRs, and we add these to SELDnet's training split for a total of up to 29 rooms (only 18 if not using \texttt{SpatialScaper}).
This process adds acoustic diversity to SELDnet's training data.
Consistent with the ``DCASE'' dataset, we generate one-minute-long  soundscapes (150 per added room), and use the same sound categories (sourced from the FSD50K dataset \cite{fonseca2021fsd50k}; our ``music'' tracks come from the FMA dataset \cite{defferrard2016fma}).
We evaluate model performance on test (i.e. STARSS ``dev-test-sony''). 

\subsection{Exp 2: Replicating ``DCASE'' with augmentations}

We also use \texttt{SpatialScaper} to recreate the ``DCASE'' dataset ``from scratch'' using real or synthetic  RIRs. We refer to these datasets as ``R real'' and ``R virt'', respectively.

To showcase \texttt{SpatialScaper}'s ability to apply effects, we create a ``R pitch'' dataset where sound events are randomly pitch-shifted half an octave (up or down) before being spatialized. 
Furthermore, we apply the channel swapping augmentation to the ``R real'' dataset, resulting in the ``R swap'' dataset.
We train separate SELDnet models using the generated data in either the ``DCASE'', ``R real'', ``R pitch'', ``R swap'', or ``R virt'' datasets.
The use of the STARSS23 dataset splits remains consistent with Exp. 1. 

\vspace{-0.3cm}
\section{Results}

\begin{figure}

\includegraphics[width=8.0cm]{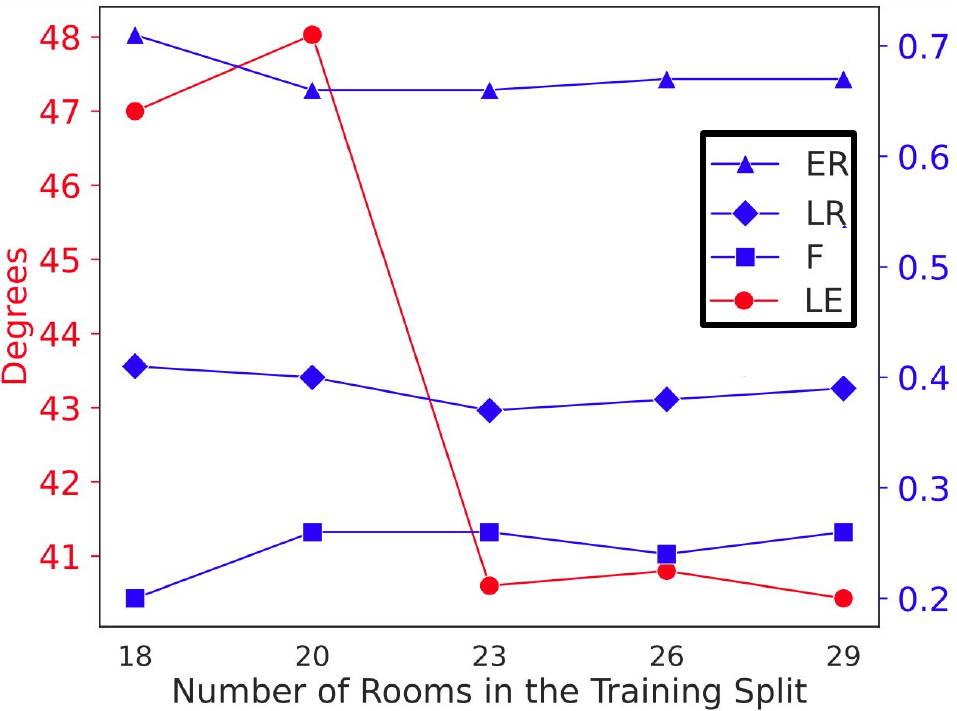}
\caption{Performance on the test split (STARSS23 ``dev-test-sony'') as a function of adding rooms (i.e. increasing acoustic diversity) to the training split.}
\label{results}
\end{figure}

Figure \ref{results} shows the results of Exp. 1. 
Adding acoustic diversity to the training data improves SELDnet's LE. 
Other metrics such as ER, F, and LR are not significantly affected.
This makes sense, since LE is determined by a model's ability to differentiate between a sound's direct path versus early wall reflections. 
Adding rooms increases the distribution of sound trajectories the model sees during training. 
Interestingly, the improved LE is observed after adding five rooms and seems to plateau.
Also note that adding only two room impacted LE, which can be explained by the fact that all new rooms come from domains and collection procedures different than the original DCASE rooms.
However, the benefit of adding rooms is unquestionable after more rooms are added. 

Something important to highlight is that SELDnet's performance before adding rooms (i.e. with its original data and training procedure) is worse than it was reported when it was released \cite{shimada2023starss23}. 
This is explained by the fact that we divided the STARSS23 ``dev-test'' split into ``dev-test-tau'' for cross-validation and ``dev-test-sony'' for testing. 
Therefore, we cross-validated SELDnet with a smaller dataset, and evaluated it on what seems to be a challenging data split\footnote{We could replicate the results in \cite{shimada2023starss23} by using all ``dev-test'' files for cross-validation and testing, as written in the code implementation by the authors.}.

\begin{table}
\begin{center}    
    \begin{tabular}{l|cccc}                  
          Data & $ER$ & $F$ & $LE$ & $LR$ \\%& $ER$ & $F$ & $LE$ & $LR$\\
         \hline
            DCASE               & 0.71 & 20.2 & 47.0 & 40.5 \\%& 0.65 & 22.7 & 51.6 & 34.3\\
            R real              & 0.71 & 19.8 & 46.5 & 34.0 \\%& 0.67 & 19.6 & 40.1 & 45.3\\
            R swap              & 0.59 & 31.7 & 29.5 & 31.2 \\%& 0.72 & 19.6 & 35.9 & 34.4\\
            R pitch             & 0.67 & 17.8 & 48.4 & 36.9 \\%& 0.72 & 16.2 & 38.8 & 37.2\\      
            R virt              & 0.75 & 7.4  & 96.7 & 19.1 \\%& 0.71 & 12.0 &	78.6 & 26.9\\
    \end{tabular}
    \caption{SELDnet performance on the STARSS23 ``dev-test-sony`` (our test split) for different versions of training data used.}
    \label{tab:resutls}
    \vspace{-18pt}
\end{center}
\end{table}

Table \ref{tab:resutls} shows results for Exp. 2. 
Comparing the first (``DCASE'') and second (``R real'') rows, we observe that  \texttt{SpatialScaper} reproduces the ``DCASE'' dataset. 
A noticeable difference is the worse LR score.
This can be attributed to the fact that we reproduced RIR databases and sound sources used in the original ``DCASE'' data, but authors did not share the music files they used. 
This is why we used the FMA dataset, which introduced a shift in the data distribution and could be the cause for the differences that we observe. 
It is worth noting that these differences are only observed in the localization metrics (LE and LR), likely due to the fact that changing the music in a room can have drastic changes in the resulting soundscape  reverberation patterns. 

We also see that using \texttt{SpatialScaper} to apply the channel-swapping augmentation leads to improvements across all metrics (except LR), an effect already described \cite{wang2023four}.
Furthermore, we observe the pitch-shift effect seems to have minimal effects. 
Finally, training with the simulated ``R virt'' data, which uses virtual RIRs, greatly impacts all metrics. 
This is not surprising, as the real RIRs better capture the reverberation properties observed in the test set data, which was collected in real rooms with human actors. 
This effect has been recently reported in the literature \cite{ickroomsim}.

\vspace{-0.2cm}
\section{Conclusions}

The presented work underscores the significance of acoustic diversity in training SELD models. 
\texttt{SpatialScaper} is an open-source library for parametric simulation and augmentation of SELD data at scale. 
Its capability to emulating varied acoustic environments proves valuable for model robustness and experimentation. 
We encourage the  broader community to explore, use, and contribute to \texttt{SpatialScaper}, fostering advancements in the SELD domain.

\section{Acknowledgements}
This work is supported by the National Science Foundation grant no. IIS-1955357. The authors thank the funding source and their grant collaborators.

\bibliographystyle{IEEEbib}
\let\oldbibitem\bibitem
\def\bibitem{\vspace{-0.5ex}\oldbibitem}
\bibliography{refs, strings}
\end{document}